\documentclass{article}
\pdfoutput=1
    \usepackage[preprint]{neurips_2019}

\usepackage[utf8]{inputenc} % allow utf-8 input
\usepackage[T1]{fontenc}    % use 8-bit T1 fonts

\usepackage{booktabs}       % professional-quality tables
\usepackage{amsfonts}       % blackboard math symbols
\usepackage{nicefrac}       % compact symbols for 1/2, etc.
\usepackage{microtype}      % microtypography
\usepackage{times}
\usepackage{soul}
\usepackage[small]{caption}
\usepackage{graphicx}
\usepackage{amsmath,amssymb,amsthm}
\usepackage{subcaption}
\usepackage{booktabs}
\usepackage[linesnumbered,lined,boxed,ruled,commentsnumbered,noend]{algorithm2e}
\usepackage{tikz}
\usepackage{xcolor}
\usepackage{thmtools}
\usepackage{natbib}

\setcitestyle{numbers,square}

\newtheorem{lemma}{Lemma}
\newtheorem{claim}{Claim}

\newtheorem{definition}{Definition}
\newtheorem{corollary}{Corollary}

\newtheorem*{theorem*}{Theorem}
\newtheorem*{lemma*}{Lemma}
\newtheorem*{claim*}{Claim}
\newtheorem*{proposition*}{Proposition}
\newtheorem*{definition*}{Definition}
\newtheorem*{corollary*}{Corollary}

\usetikzlibrary{arrows}

\newcommand{\cA}{{\cal A}}

\newcommand{\cC}{{\cal C}}
\newcommand{\cD}{{\cal D}}

\newcommand{\cE}{{\cal E}}

\newcommand{\cH}{{\cal H}}

\newcommand{\cK}{{\cal K}}

\newcommand{\cO}{{\cal O}}
\newcommand{\cP}{{\cal P}}

\newcommand{\cR}{{\cal R}}

\newcommand{\cT}{{\cal T}}

\newcommand{\origin}{s^{\star}}
\newcommand{\NP}{{\mathcal NP}}
\newcommand{\deja}{D{\'e}j{\`a} Vu }

\declaretheorem{theorem} 
\declaretheoremstyle[%
  spaceabove=-6pt,%
  spacebelow=6pt,%
  headfont=\normalfont\itshape,%
  postheadspace=1em,%
  qed=\qedsymbol%
]{mystyle} 
\declaretheorem[name={Proof},style=mystyle,unnumbered,
]{prf}

\title{On the Constrained Least-cost Tour Problem}

\author{%
  Patrick O'Hara \\
  The Alan Turing Institute \\
  \small{{\tt pohara@turing.ac.uk}} \\
  \AND
  M.S. Ramanujan \\
  Dept. of Computer Science \\
  University of Warwick \\
  \small{{\tt r.maadapuzhi-sridharan@warwick.ac.uk}} \\
  \AND
  Theodoros Damoulas \\
  The Alan Turing Institute \\
  Depts. of Computer Science and Statistics \\
  University of Warwick \\
  \small{\texttt{t.damoulas@warwick.ac.uk}} \\
}

\begin{document}

\maketitle

\begin{abstract}
  We introduce the Constrained Least-cost Tour (CLT) problem:
  given an undirected graph with weight and cost functions on the edges, minimise the total cost of a tour rooted at a start vertex such that the total weight lies within a given range.
  CLT is related to the family of Travelling Salesman Problems with Profits, but differs by defining the weight function on edges instead of vertices, and by requiring the total weight to be within a range instead of being at least some quota.
  We prove CLT is $\mathcal{NP}$-hard, even in the simple case when the input graph is a path.
  We derive an informative lower bound by relaxing the integrality of edges and propose a heuristic motivated by this relaxation.
  For the case that requires the tour to be a simple cycle, we develop two heuristics which exploit Suurballe's algorithm to find low-cost, weight-feasible cycles.
  We demonstrate our algorithms by addressing a real-world problem that affects urban populations: finding routes that minimise air pollution exposure for walking, running and cycling in the city of London.
\end{abstract}

\section{Introduction}

Large scale graph optimisation problems are at the heart of urban science, computational sustainability \cite{gomes2009computational} and human wellbeing.
Examples include computing connected subgraphs to design wildlife corridors \cite{Dilkina2010} and planning methods for bike sharing systems in New York \cite{Freund2019}.
With $91\%$ of the world population exposed to particulate matter (PM2.5) concentrations that are above the annual mean World Health Organization air quality guideline levels \cite{WHO2016}, we are motivated by the interesting and challenging problem of finding running routes that minimise air pollution in a city.

Consider a runner planning a route which starts and ends at the same location. 
The runner would like the route to be sufficiently long but they do not want to run too far.
However, running without considering the air quality in the local area is a suboptimal approach. 
Air pollution has an adverse effect on the cardio-respiratory system, which can be exacerbated by increased inhalation during exercise \cite{Giles2014}. 
Moreover, air pollution in urban environments is highly localised because factors such as transportation, industry and construction largely contribute to the poor air quality \cite{Vardoulakis2008}. 
In order to minimise the exposure to air pollution, the runner could use a mobile or web application to request and view an appropriate route. 
An algorithm which computes such a running route efficiently is therefore highly desirable.

% motivation inspires problem defintion

With the above motivation in mind, we study the Constrained Least-cost Tour (CLT) problem. 
The input is an undirected graph representing the road network with edge weights (distance) and edge costs (air pollution).
We are given a lower weight threshold $W_1$ and an upper weight threshold $W_2$.
We are also given a specially annotated vertex called the origin representing the start and end location of the run.
The objective is to minimise the total cost of a tour starting and ending at the origin such that the total weight of the tour is weight-feasible. 
A {\em weight-feasible} tour means that the total weight of the tour is at least $W_1$ and at most $W_2$.
In the context of a running route, weight-feasible means that the route is sufficiently long (at least $W_1$) but is not too far for the runner (at most $W_2$).

% literature related to CLT

\begin{table}[t]
  \caption{Summary of CLT and {\sc TSP-wP}s. Cost on edges $c:E(G) \to \mathbb{N}$; weight on edges $w:E(G) \to \mathbb{N}$; profit on vertices $p:V(G) \to \mathbb{N}$; resource on vertices $r:V(G) \to \mathbb{N}$; threshold $Z \in \mathbb{N}$.}
  \label{tab:lit}
\centering
\begin{tabular}{l c c c c l r}
  \toprule
  Problem & $c$ & $w$ &  $p$ & $r$ & Objective & Constraints \\
  \midrule
  CLT & yes & yes & no & no & $\min\{c(\cT)\}$ & $W_1 \leq w(\cT) \leq W_2$ \\
  OP & yes & no & yes & no & $\max\{p(\cT)\}$ & $c(\cT) \leq Z$ \\
  {\sc Pc-TSP} & yes & no & yes & yes & $\min\{c(\cT) - p(\cT)\}$ & $r(\cT) \geq Z$ \\
  {\sc Q-TSP} & yes & no & yes & no & $\min\{c(\cT)\}$ & $p(\cT) \geq Z$ \\
  \bottomrule
\end{tabular}
\end{table}

The CLT problem is most closely related to the family of Travelling Salesman Problems with Profits ({\sc TSP-wP}). Table~\ref{tab:lit} summarises the similarities and differences between CLT and some {\sc TSP-wP} family members.
This family of TSPs does not require the tour to visit every vertex in the graph and may be rooted (tour must start and end at a given vertex) or unrooted.
In a review by Feillet et al. \cite{Feillet2005}, the {\sc TSP-wP} family is split into three classes: Quota TSP ({\sc Q-TSP}), Selective TSP ({\sc S-TSP}) and Profitable Tour Problems (PTP). The Orienteering Problem (OP) \cite{Golden1987,Gunawan2016} is a well-known {\sc S-TSP} and the Prize-collecting TSP ({\sc Pc-TSP}) \cite{Balas1989} is an important {\sc PTP}.
However, the more similar {\sc TSP-wP} class to the CLT probem is {\sc Q-TSP} \cite{Awerbuch1995}.

% the Quota TSP

In {\sc Q-TSP}, we are given an undirected graph with a profit function on the vertices, a cost function on the edges and a quota $Z \in {\mathbb N}$.
The goal of {\sc Q-TSP} is to minimise the total cost of a tour such that the total collected profit is at least $Z$.
There are three key differences between the CLT Problem and the {\sc Q-TSP}.
First, the profit function in {\sc Q-TSP} is on the vertices of the graph whereas the weight function in the CLT Problem is on the edges of the graph.
Second, there is no upper profit threshold in {\sc Q-TSP}, where as the CLT Problem defines an upper weight threshold $W_2$.
Third, the limited existing literature \cite{Awerbuch1995} for {\sc Q-TSP} assumes the triangle inequality holds on the cost function and obtains an approximation by doubling a $k$-Minimum Spanning Tree \cite{Garg2005}. 
However, the cost function (air pollution exposure) in our real-world application does not follow this inequality.

% two variants of CLT and NP-hardness

{\sc TSP-wP}s such as Q-TSP may ask for a {\em weak} tour (vertex repetition is not constrained) or a {\em strong} tour (vertices are repeated at most once, thus the tour is a simple cycle).
We investigate both the weak and strong variants of the CLT problem.
In the remainder of this paper, we refer to the weak variant as the CLT problem, and the strong variant as the Constrained Least-cost Cycle (CLC) Problem.

% contributions and overview of paper

Our contributions are as follows. 
In Section \ref{sec:complexity}, we prove that the CLT problem is $\mathcal{NP}$-hard, even in the case when the input graph is a path. 
A simple reduction from the Hamiltonian Cycle problem shows that if the triangle inequality does not hold on the cost function, then the CLC problem does not have an $\alpha$-approximation algorithm for any $\alpha \geq 1$ (assuming $\cP \neq \NP$).
Thus, we will focus on heuristic approaches to the problem which find sufficiently good solutions in polynomial-time (Section \ref{sec:heuristic}).
First, we introduce the \deja heuristic (DjV) to find weak tours for the CLT problem.
Next, we propose Suurballe's heuristic (SH), which calls on Suurballe's algorithm \cite{suurballe1984} to find solutions to the CLC problem.
Finally, we develop the Adaptive heuristic (AH) that extends SH by exploring a greater proportion of the solution space.
We analyse the performance of our heuristics on two datasets by comparing them against the continuous and connectivity relaxations of the CLT and CLC problems respectively. 

\section{Problem definition} \label{sec:definition}

\begin{definition}
  A tour $\mathcal{T} = (v_1, \ldots, v_n)$ is a sequence of vertices starting and ending at the same vertex, where every two consecutive vertices in the sequence are adjacent to each other in the graph.
\end{definition}

In the CLT problem, we are given an undirected graph $G$ where $V(G)$ denotes the set of vertices and $E(G)$ denotes the set of edges. Each edge has a weight function defined by $w : E(G) \to \mathbb{N}$ and a cost function defined by $c : E(G) \to \mathbb{N}$. We are also given a vertex $s^{\star} \in V(G)$ and weight thresholds $W_1,W_2 \in {\mathbb N}$ where $W_1 \leq W_2$. 
Let $\phi(\mathcal{T}, e)$ denote the number of times edge $e \in E(G)$ is repeated in tour $\mathcal{T}$.
We denote the total cost and weight of a tour $\mathcal{T}$ by $c(\mathcal{T}) = \sum_{e \in \mathcal{T}}~{\phi(\mathcal{T}, e) \cdot c(e)}$ and $w(\mathcal{T}) = \sum_{e \in \mathcal{T}}~{\phi(\mathcal{T}, e) \cdot w(e)}$.
We say a tour $\cT$ is {\em weight-feasible} if and only if $W_1 \leq w(\cT) \leq W_2$.

The goal of the CLT problem is to minimise the total cost of a tour starting and ending at $s^{\star}$ such that the tour is weight-feasible. 
That is, the goal is to minimise $c(\mathcal{T})$ subject to $W_1 \leq w(\mathcal{T}) \leq W_2$ and $s^{\star} = v_1 = v_n$.
In the {\sc CLT Decision} problem, we ask if there is a tour $\mathcal{T} = (v_1, \ldots, v_n)$ such that $W_1 \leq w(\mathcal{T}) \leq W_2$ and $c(\mathcal{T}) \leq C$ where $C \in {\mathbb N}$. 
Finally, we define the Constrained Least-cost Cycle (CLC) problem, which has the same objective function and constraints as CLT, but adds the requirement that  
the tour must be a simple cycle that visits vertices at most once. 

\section{$\mathcal{NP}$-hardness} \label{sec:complexity}

We prove that {\sc CLT Decision} is $\mathcal{NP}$-hard by reducing from the unbounded subset sum problem, which is known to be $\mathcal{NP}$-hard~\cite{Karp1972,Kellerer2004}. For brevity, we continue to refer to this variant as {\sc Subset Sum}. We construct an instance of {\sc CLT Decision} on a path, starting from an instance of {\sc Subset Sum}. 
We show that an instance of {\sc Subset Sum} is a {\sc Yes}-instance if and only if our constructed instance of {\sc CLT Decision} is a  {\sc Yes}-instance.

\begin{definition}
In the {\sc Subset Sum} problem, we are given a set of $n$ items ${\cal X} = \{z_1,\dots, z_n\}$, a target value $Z\in {\mathbb N}$ and weight function $f:{\cal X}\to {\mathbb N}$,
and want to decide whether there is a vector $(y_1,\dots, y_n)\in ({\mathbb N_{\geq 0}})^n$ 
such that $Z=\sum_{i \in [n]}~y_i\cdot f(z_i)$.
\end{definition}
\begin{theorem} \label{thm:np-hard}
The Constrained Least-cost Tour problem is $\mathcal{NP}$-hard, even if the input graph is a path.
\end{theorem}
\begin{prf}Let $I=({\cal X}=\{z_1,\dots, z_n\},Z,f)$ be an instance of {\sc Subset Sum}.
  We construct an instance of {\sc CLT Decision} as follows. 
  Let $G$ denote the graph on vertex set $v_1,\dots, v_{n+2}$ and edge set $\{e_1,\dots, e_{n+1}\}$ where $e_i$ has endpoints $v_i$ and $v_{i+1}$ for every $i\in [n+1]$. 
  In other words, $G$ is a path on $n+2$ vertices. 
  For all $i\in [n]$, we define the weight function $w:E(G)\to {\mathbb N}$ as $w(e_i)=f(z_i)$ and the cost function $c:E(G)\to {\mathbb N}$ as $c(e_i)=f(z_i)$.
  For the last edge $e_{n+1}$, we set $w(e_{n+1})$ to be $(4Z+4\sum_{i\in [n]}w(e_i))^2$ and $c(e_{n+1})$ to be $\frac{1}{2} \cdot (2Z+2\sum_{i\in [n]}w(e_i))$.
  We set $C$ to be $4Z+4\sum_{i\in [n]}w(e_i)$, $W_1$ to be $2Z+2w(e_{n+1})+2\sum_{i\in [n]}w(e_i)$ and $W_2$ to be any positive integer such that $W_1 \leq W_2$.
  Finally, we assign $s^{\star}=v_1$. 
  This completes the construction of the {\sc CLT Decision} instance.
  Clearly, the reduction is polynomial-time.  We now argue that $I$ is a {\sc Yes}-instance of {\sc Subset Sum} if and only if the constructed instance is a {\sc Yes}-instance of {\sc CLT Decision}.
  
  Suppose that $I$ is a {\sc Yes}-instance of {\sc Subset Sum} and let $(y_1,\dots, y_n)$ be the solution. Consider the ``natural" tour $\mathcal{T}$ in $G$ starting at $v_1$, traversing every $e_i$ ($i\in [n]$) exactly $2y_i+2$ times and the edge $e_{n+1}$ exactly twice. Then, the values of $w(\mathcal{T})$ and $c(\mathcal{T})$ are precisely the following:
  \begin{align*}
  w(\mathcal{T}) &= \sum_{i\in [n]}~w(e_i)(2y_i+2)+2w(e_{n+1})=2\overbrace{\sum_{i\in [n]}~{y_i f(z_i)}}^{Z}+2\sum_{i\in[n]}~w(e_i) + 2 w(e_{n+1})=W_1. \\
  c(\mathcal{T}) & = \sum_{i\in [n]}~c(e_i)(2y_i+2)+2c(e_{n+1})=2\overbrace{\sum_{i\in [n]}~y_i\cdot f(z_i)}^{Z}+4\sum_{i\in[n]}~w(e_i) + 2Z =C
  \end{align*}
  Hence, we conclude that if $I$ is a {\sc Yes}-instance, then the constructed instance of {\sc CLT Decision} is a {\sc Yes}-instance. 
  Conversely, suppose that the constructed instance of {\sc CLT Decision} is a {\sc Yes} instance. This implies a tour $\mathcal{T}$ in $G$ starting at $v_1$ such that $W_1 \leq w(\mathcal{T})\leq W_2$ and $c(\mathcal{T})\leq C$.
  
  \begin{claim}
    $\mathcal{T}$ traverses every edge of $G$ at least twice and the edge $e_{n+1}$ exactly twice.
  \end{claim}
  
  \begin{prf}
    Observe that in order to prove the claim, it is sufficient to prove that $\mathcal{T}$ traverses $e_{n+1}$ {\em exactly twice}.
    $\cT$ must traverse $e_{n+1}$ {\em at least} twice, since $e_{n+1}$ cannot be traversed once, and not traversing $e_{n+1}$ would contradict our assumption that $c(\mathcal{T})\leq C$.
    $\mathcal{T}$ must traverse $e_{n+1}$ {\em at most} twice, since otherwise $\cT$ traverses $e_{n+1}$ at least four times, and we obtain $c(\cT) > 4 \cdot c(e_{n+1})$ and $c(\cT) \leq C$ which is a contradiction. 
  \end{prf}
  
  We now describe the solution vector $(y_1,\dots, y_n)$ for the {\sc Subset Sum} instance.
  Recall we have already proved that for all $e\in E(G)$, $\phi(\mathcal{T},e)\geq 2$, $\phi(\mathcal{T},e)$ is even, and $\phi(\mathcal{T},e_{n+1})=2$.
  For every $i\in [n]$, define $y_i=\frac{1}{2} \phi(\mathcal{T},e_i)-2.$ Clearly, $y_i\in {\mathbb N}\cup \{0\}$ for all $i\in [n]$ as required in the description of {\sc Subset Sum} and since  $\mathcal{T}$ traverses $e_{n+1}$ exactly twice, we conclude that $\sum_{i\in [n]} y_i\cdot w(z_i)=Z$.
  Here, we use the fact that $w(\mathcal{T})\geq W_1$ to imply that $\sum_{i\in [n]} y_i\cdot w(z_i)\geq Z$ and use $c(\mathcal{T})\leq C$ to infer that $\sum_{i\in [n]} y_i\cdot w(z_i)\leq Z$.
  This completes the proof in the converse direction.
\end{prf}

Theorem~\ref{thm:np-hard} proves CLT is $\NP$-hard when the input graph is a path, so clearly CLT is $\NP$-hard on a general graph.
The CLC problem can also be shown to be $\NP$-hard by reducing from the Hamiltonian Cycle problem.
We further note that if the cost function does not satisfy the triangle inequality, then CLC does not have an $\alpha$-approximation algorithm
for any $\alpha \geq 1$ (assuming $\cP \neq \NP$).
\footnote{See appendix for complete proofs.}

\section{Relaxations} \label{sec:relaxation}

Relaxing $\mathcal{NP}$-hard problems often provides useful insights about the optimal solution to the original problem and provides a lower bound we can compare our heuristics against. 
In this work, we consider two relaxations. 
Firstly, we show that on continuous graphs, the CLT problem has a polynomial-time algorithm that finds the optimal solution. 
In a {\em continuous graph} each edge is viewed as infinitely many vertices of degree two with infinitesimally small edges (formally the continuous graph is the geometric realisation of the graph topology). 
This is equivalent to saying that the multiplicity of an edge can be any positive real value ($\phi(\mathcal{T}, e) \in {\mathbb R}^+$). 
Secondly, we give an integer programming formulation of CLC and relax the constraint that the solution must be connected.

\subsection{Continuous relaxation}

Let $G_c$ denote a continuous graph. 
We define the induced graph $H_c$ over a tour $\mathcal{T}_c$ of $G_c$ as follows. 
A vertex $v$ is in $H_c$ if $v \in \mathcal{T}_c$. An edge $e$ is in $H_c$ if the multiplicity of the edge in $\mathcal{T}_c$ is greater than zero. 
If $H_c$ is a path $v_1,\dots, v_n$ where $v_1$ is the origin and $e_i=(v_i,v_{i+1})$ for each $i\in [n-1]$,  then we call the edge $e_{n-1}$ the \textit{head} of the tour, and the remaining edges the \textit{tail} of the tour.
We argue that in this relaxation, we may assume that the optimal tour induces a path in $G_c$.   

\begin{lemma} \label{lem:continuous}
On a continuous graph, the induced graph of the optimal tour is a path.
\end{lemma}
\begin{prf}
Take an optimal tour $\cT^*_c$ and let $e=(u,v)$ be an edge in the induced graph $H_c^*$ which has the least cost per unit weight among the edges in $H_c^*$. 
Without loss of generality, suppose that $u$ is the first vertex of $e$ which is visited by $\cT^*_c$ in the traversal starting from $\origin$. 
Let $P$ denote a minimum cost path in $G_c$ from $s^\star$ to $u$. On the one hand, $w(P)$ may already be greater than $w(\cT^*_c)$, in which case we can simply find another tour contained within $P$, of same weight as $w(\cT^*_c)$ and at most the same cost as $c(P)$, which in turn must be at most $c(\cT^*_c)$. 
But, $w(P)$ may be less than $w(\cT^*_c)$. However, the cost of any subtour of $\cT^*_c$ which starts at $\origin$ and visits $u$ must still have cost at least that of $P$. 

Now, consider the tour obtained by starting at $\origin$, traversing $P$ and arriving at $u$, followed by taking $e$ with multiplicity  $\frac{1}{w(e)}(w(\cT^*_c)-2w(P))$, followed by taking $P$ all the way back to $\origin$. Call this tour $\cD^*_c$. By definition, $w(\cD^*_c)=w(\cT^*_C)$. Moreover, it is straightforward to see that $c(\cD^*_c)\leq c(\cT^*_c)$. 
\end{prf}

Thus, we can compute an optimal solution to the continuous relaxation in polynomial-time by running Dijkstra's algorithm \cite{dijkstra1959}, computing the multiplicity of the head for every edge $e \in G_c$, and returning the least-cost solution. 
We also have a lower bound on the cost of an optimal {\em discrete} solution:

\begin{corollary}
The cost of the optimal constrained least-cost tour $\mathcal{T}_c^*$ on graph $G_c$ is less than or equal to the cost of the optimal constrained least-cost tour $\mathcal{T}$ on $G$.
\end{corollary}

\subsection{Connectivity relaxation}

We first turn our attention to an integer programming (IP) formulation for the CLC problem. 
Let $e_{ij} \in E(G)$ be an edge between vertex $i$ and $j$. 
We place the variable $x_{ij} \in \{0,1\}$ on the edges of the graph. 
The objective function is to minimise $\sum_{ij}~{x_{ij} \cdot c(e_{ij})}$ subject to $W_1 \leq \sum_{ij}~{x_{ij} \cdot w(e_{ij})} \leq W_2$ for all $e_{ij} \in E(G)$.
Let $x(S) = \sum_{ij}~{x_{ij}}$ for a subset $S \in E(G)$ and $\mathcal{A}_i \subset E(G)$ be the set of edges $e_{ij}$ adjacent to vertex $i$. 
We add the constraints $x(\mathcal{A}_s) = 2$ to enforce $\mathcal{T}$ starts and ends at the origin $\origin$, and $\forall i \in V(G): \frac{1}{2} x(\mathcal{A}_i) \in \{0,1\}$ to ensure $\mathcal{T}$ is closed. 
Finally, we add the sub-tour elimination constraint \cite{Bauer2002,Laporte1986} to enforce connectivity.
Optimally solving an IP that enforces connectivity is not be practical in our application of a runner requesting a route, since the user might have to wait for too long as the size of the graph and length of the requested run increase.
Thus we relax the connectivity constraint and use the resulting IP as a lower bound on the optimal solution of CLC to compare our heuristic against.

\section{Heuristics} \label{sec:heuristic}

\begin{figure*}[t]
  \centering
  \begin{subfigure}[t]{0.32\textwidth}
    \includegraphics[width=\linewidth]{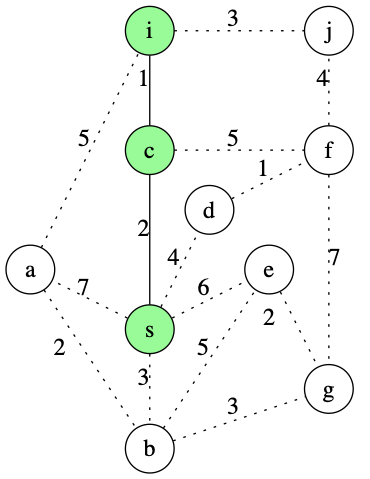}
  \end{subfigure}
  ~
  \begin{subfigure}[t]{0.32\textwidth}
    \centering
    \includegraphics[width=\linewidth]{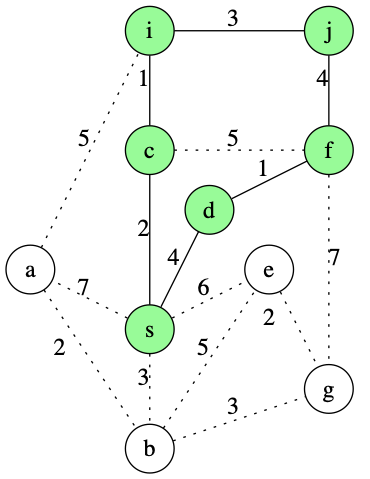}
  \end{subfigure}
  ~
    \begin{subfigure}[t]{0.32\textwidth}
    \includegraphics[width=\linewidth]{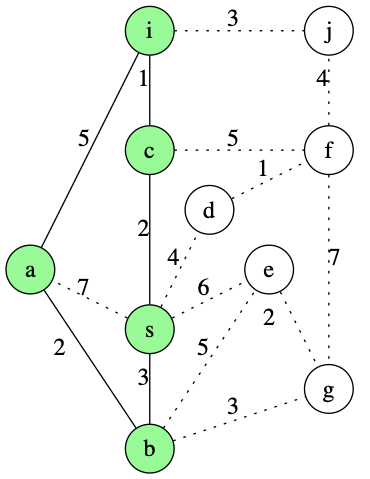}
  \end{subfigure}
  \caption{Examples of tours for each heuristic. We are given an undirected graph $G$ with unit weights and costs labelled on edges. We ask for a tour which minimises the total cost starting at $s$ such that $5 \leq w(\cT) \leq 7$. Green vertices and solid edges are in the tour. White vertices and dotted edges are not in the tour. [{\em Left}] DjV where $\phi(\cT, e_{c,i}) = 4$ and $\phi(\cT, e_{s,c}) = 2$ such that $c(\cT) = 8$ and $w(\cT) = 6$. [{\em Middle}] SH finds a simple cycle with $c(\cT) = 15$ and $w(\cT) = 6$. [{\em Right}] AH computes a simple cycle with $c(\cT) = 13$ and $w(\cT) = 5$.}
  \label{fig:graph-examples}
\end{figure*}

Our approach is to develop heuristics that run in polynomial-time and return close to optimal solutions in real-world environments.
Fig.~\ref{fig:graph-examples} shows examples for each algorithm.
First, we present the \deja (DjV) heuristic for the CLT problem that exploits the continuous relaxation.\footnote{The repetitive nature of the low-cost edge is where the name ``D{\'e}j{\`a} Vu'' comes from.}
Next, we propose Suurballe's heuristic (SH) that finds low-cost cycles for the CLC problem. Finally, we introduce the Adaptive heuristic (AH) which extends SH to explore a greater proportion of the solution space.

\subsection{\deja Heuristic}

Our \deja (DjV) heuristic exploits the intuition that the optimal solution to CLT on a continuous graph is a good indicator of a low-cost solution to CLT on a discrete graph. 
DjV walks along a path to an edge with low cost, repeats this edge with some positive {\em even} multiplicity, then walks back along the same path to the origin.

DjV computes the least-cost tree rooted at $\origin$. 
We store the parent $\pi_v$ and cost $l_v$ of the least-cost path $\Pi(\origin,v)$ from $\origin$ to $v$.
For each edge $e \in E(G)$, we extract $\Pi(\origin,x)$ where $x$ is the endpoint of $e$ with least $l_x$. 
If $2 w(\Pi(\origin,x)) + 2 w(e) \leq W_2$, then the multiplicity of edge $e$ in tour $\mathcal{T}$ is 
$$\phi(\cT, e) = f\left(\frac{W_1 - 2 w(\Pi(\origin,x))}{w(e)}\right)$$
where $f$ rounds up to the closest positive even integer.
The time-complexity of DjV is $\cO(n \log n)$.

\subsection{Suurballe's heuristic}

We now propose Suurballe's heuristic (SH) which uses the fact that a pair of vertex-disjoint simple paths between two vertices $u$ and $v$ form a simple cycle $\cC$.
Suurballe's algorithm \cite{Suurballe1974,suurballe1984} solves the {\sc Shortest Pairs} of disjoint paths problem:
given a directed, weighted graph $G^\prime$, find a pair of edge-disjoint paths with minimum total cost from a source vertex $\origin$ to a sink vertex $v$, for every possible sink $v \in G^\prime$.
Suurballe and Tarjan \cite{suurballe1984} give an algorithm for {\sc Shortest Pairs} with time complexity $\cO(m \log n)$. 
Their algorithm requires $G^\prime$ to be asymmetric, that is if $(x,y)$ is an arc in $G^\prime$, then $(y,x)$ is not in $G^\prime$.
To construct a directed, asymmetric graph $G^\prime$ from our undirected graph $G$, we use the vertex splitting transformation as described by Suurballe and Tarjan. 
The splitting transformation also allows us to compute vertex-disjoint paths on $G^\prime$ using Suurballe and Tarjan's edge-disjoint algorithm.

Given $G^\prime$, SH runs Suurballe's algorithm and computes the cost of the shortest pairs of vertex-disjoint paths $\Lambda(\origin, v)$ from $\origin$ to every vertex $v \in V(G^\prime)$. 
From $\Lambda(\origin, v)$, we construct a simple cycle $\cC$ containing $\origin$ and $v$. 
We return the least-cost simple cycle $\cC$ that is weight-feasible ($W_1 \leq w(\cC) \leq W_2$). 
The time complexity of the heuristic is $\cO(m \log n)$ under the assumption that the number of vertices in $\cC$ is much smaller than the number of edges in the graph. 
The weakness of SH is it only considers a small subset of the solution space to CLC. 
This subset covers all cycles which can be constructed by finding the least-cost pair of vertex-disjoint paths from $s^\star$ to a vertex $v$ in the graph.

\subsection{Adaptive heuristic}

The Adaptive heuristic (AH) extends SH by exploring a larger solution space.
This space encompasses all cycles containing the origin that are formed by computing the least-cost pair of vertex-disjoint paths between every pair of vertices in the graph.
Thus the solution space of SH is a subset of the solution space of AH.
Psuedocode for AH is given by Algorithm \ref{alg:adaptive}.
The time-complexity of AH is $\cO(n \cdot m \log n)$ because we must compute Suurballe and Tarjan's algorithm \cite{suurballe1984} at most $n = |V(G)|$ times.
The increase in time-complexity is the price to pay for exploring more solutions.

\begin{algorithm}[h]
    \DontPrintSemicolon
    \SetKwInOut{Input}{input}\SetKwInOut{Output}{output}
    \caption{Adaptive heuristic for the Constrained Least-cost Cycle Problem}
    \label{alg:adaptive}
    \Input{An undirected graph $G$ with edge weights $w:E(G) \to \mathbb{N}$ and edge costs $c:E(G) \to \mathbb{N}$; the origin $\origin \in V(G)$; lower and upper weight thresholds $W_1, W_2 \in \mathbb{N}$.}
    \SetAlgoLined

    Construct a directed, asymmetric graph $G^\prime$ from the undirected graph $G$. \;
    For every pair of vertices $u,v \in V(G^\prime)$, compute the least-cost pair of vertex-disjoint paths $\Delta(u,v)$ between $u$ and $v$. \;
    From $\Delta(u,v)$, construct a simple cycle $\cC$. If $\origin \in \cC$, then $\cC$ is a candidate solution. \;
    \Output{The least-cost weight-feasible simple cycle $\cC^\star$ from all candidate solutions $\cC$.}
\end{algorithm}

\section{Experiments} \label{sec:experiments}

We now present the results from running our heuristics on two datasets and comparing them against our relaxations.\footnote{AH = adaptive heuristic, CR = continuous relaxation, XR = connectivity relaxation, SH = Suurballe's heuristic, DjV = \deja heuristic. Code available at \url{https://patrickohara.github.io/CLT-problem/}.}
DjV is compared against CR.
SH and AH are compared against XR.
Every algorithm is tested at 10 different weight thresholds and 10 random origins. 
The gap between $W_1$ and $W_2$ is kept constant at 250 meters for the pollution dataset and 5 units for the Crucible dataset. 
Experiments for the heuristics and continuous relaxation are computed on a Microsoft Azure virtual machine with four CPUs and 14GB of memory running Linux. 
The connectivity relaxation is computed using the IBM Decision Optimisation Cloud service with 10 cores and 60GB of memory. 
We set the maximum time limit for IP to solve the connectivity relaxation to one hour.
We reduce the size of our input graph with two pre-processing steps. 
The first removes vertices which cannot be reached from $\origin$ within $W_2/2$. 
The second removes all leaves from the graph when solving the CLC problem, and so is only applied to SH, AH and XR.

\begin{figure*}[t]
  % \centering
  % \begin{subfigure}[t]{0.32\textwidth}
  %   \includegraphics[width=\textwidth]{figures/pollution_dataset}
  % \end{subfigure}
  % ~
  \begin{subfigure}[t]{0.48\textwidth}
    \includegraphics[width=\textwidth]{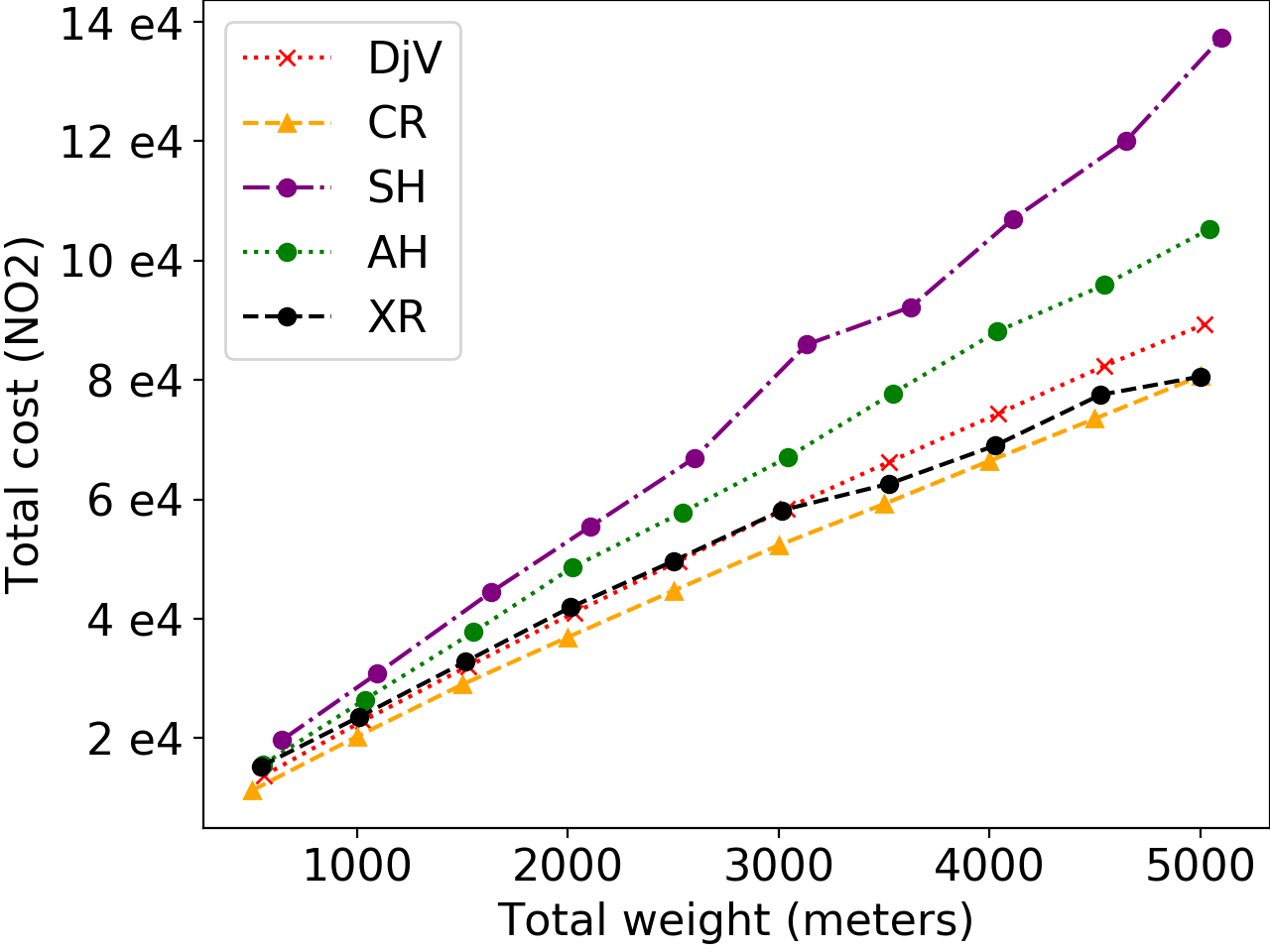}
  \end{subfigure}
  ~
  \begin{subfigure}[t]{0.48\textwidth}
    \includegraphics[width=\textwidth]{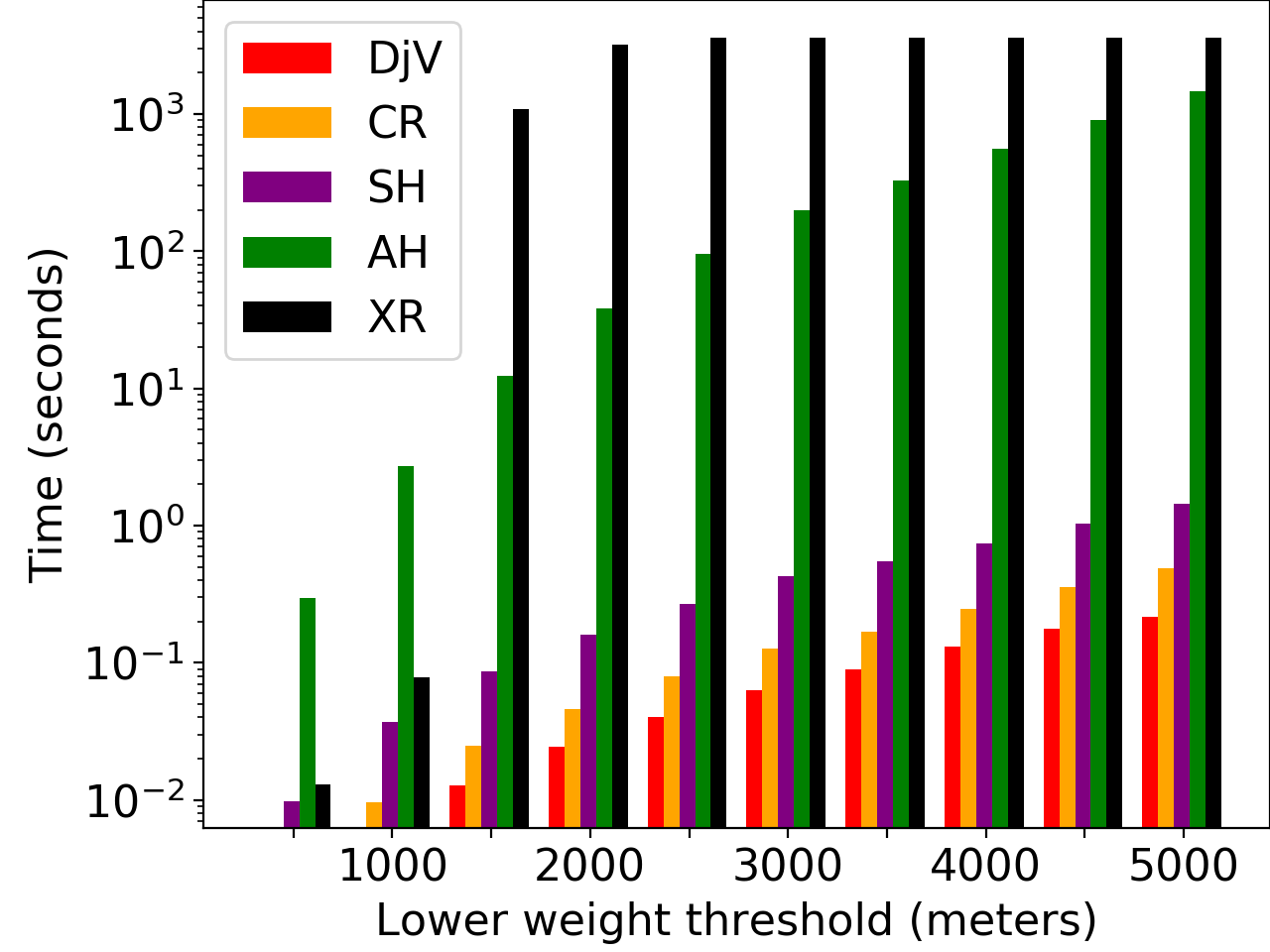}
  \end{subfigure}

  \caption{[{\em Left}] Mean total distance of running route vs mean total $NO_2$. [{\em Right}] Time taken for each algorithm to execute. See $^3$ for labels.}
  \label{fig:pollution_dataset}
\end{figure*}

\begin{figure*}[t]
  \centering
  \begin{subfigure}[t]{0.48\textwidth}
    \includegraphics[width=\textwidth]{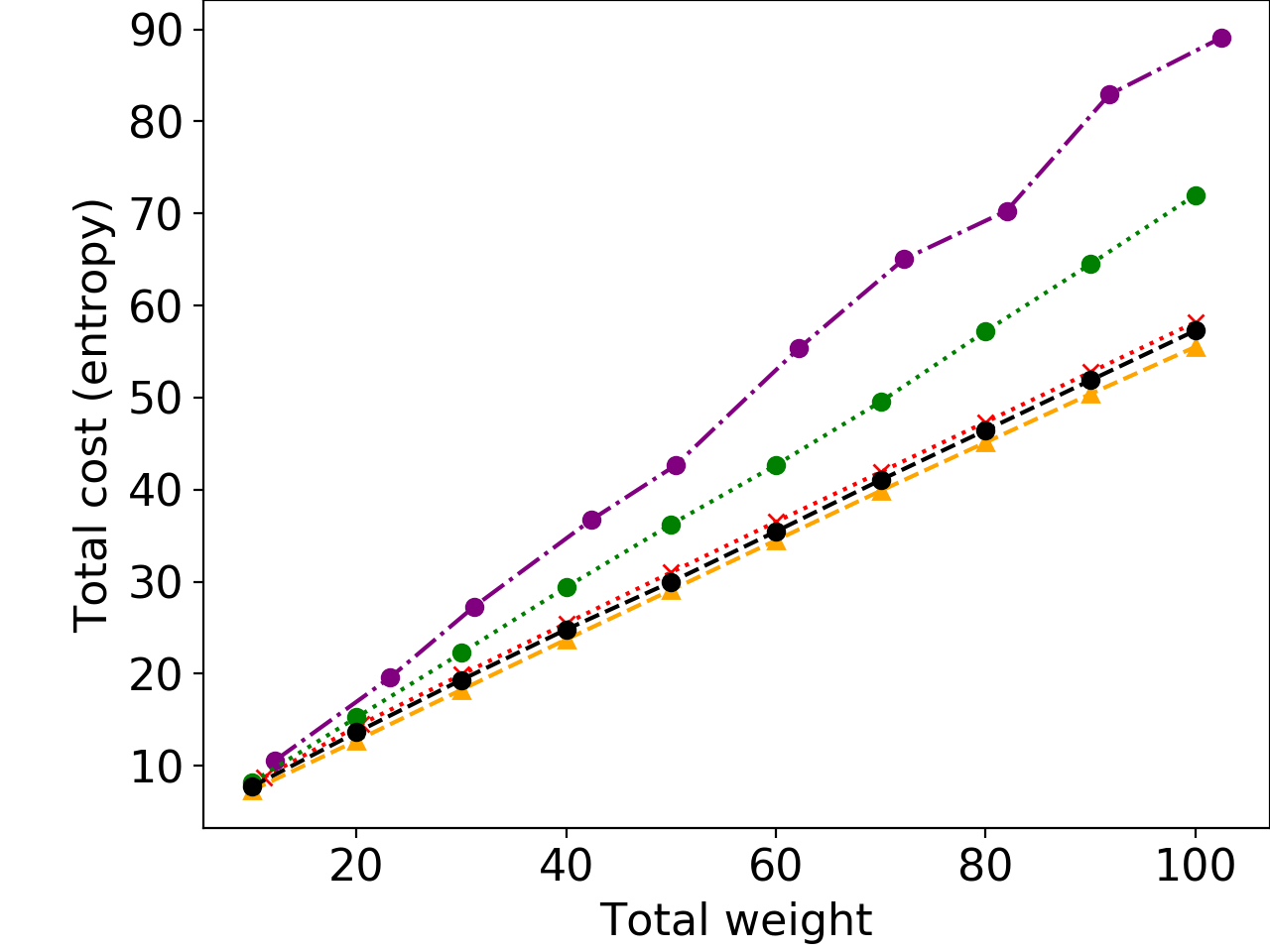}
  \end{subfigure}
  ~
    \begin{subfigure}[t]{0.48\textwidth}
        \includegraphics[width=\textwidth]{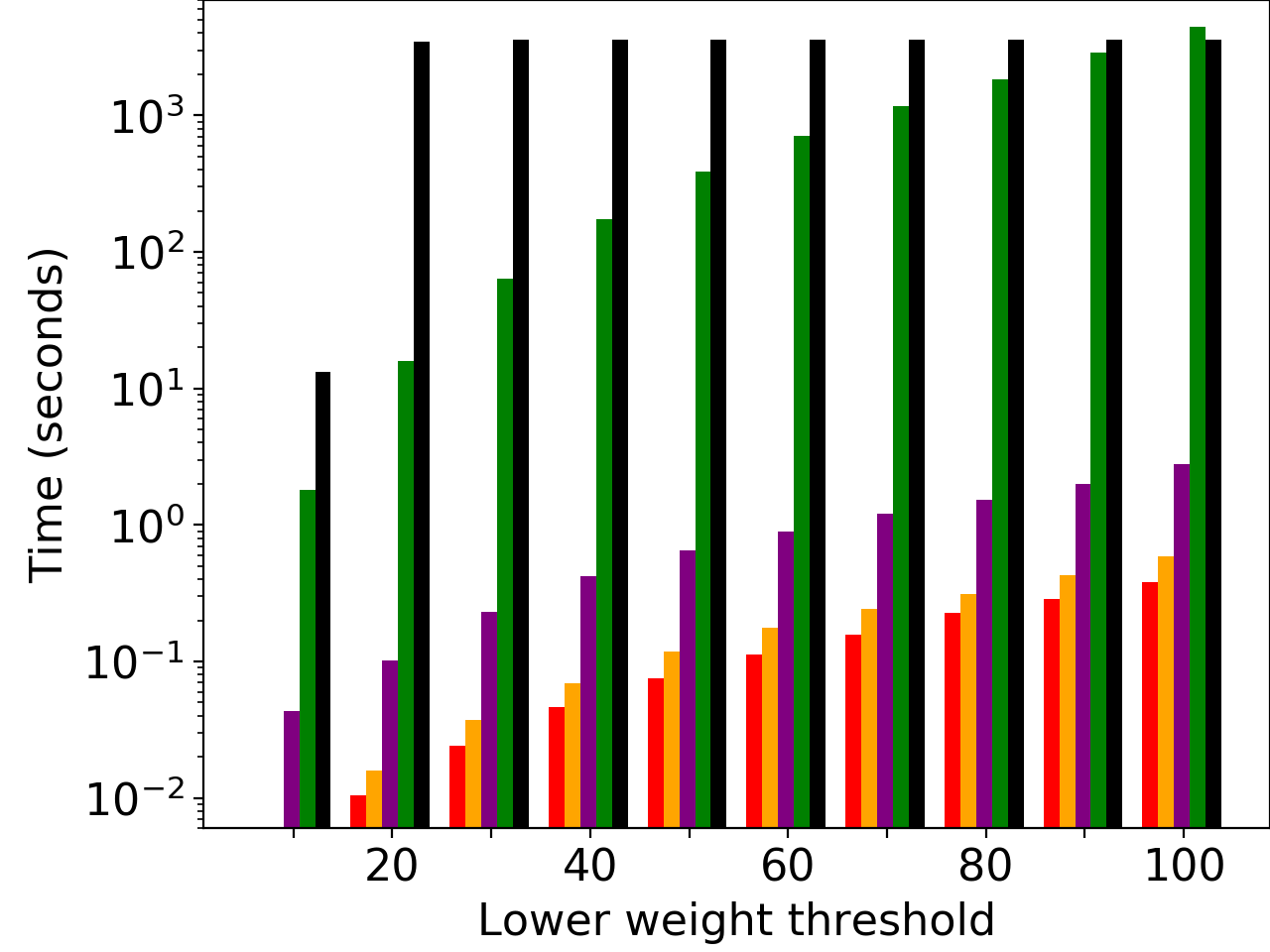}
    \end{subfigure}
    \caption{Results for the Crucible dataset. Labelling$^3$ and colours same as Fig.~\ref{fig:pollution_dataset}.
    [{\em Left}] Mean total weight vs mean total cost. 
    [{\em Right}] Time taken.}
  \label{fig:map_dataset}
\end{figure*}

Figs.~\ref{fig:pollution_dataset} and \ref{fig:map_dataset} show the effect of increasing the weight thresholds. 
Figs~\ref{fig:dejavu-crucible} and \ref{fig:examples} display examples of routes computed by our heuristics for the Crucible and air quality dataset respectively. 
Table~\ref{tab:results} summarises the overshoot and margin of error. 
The overshoot of a tour $\cT$ is defined as $w(\cT) - W_1$.
The margin of error of a heuristic $\cH$ against its relaxation $\cR$ is defined as $100 \times (c(\cH) - c(\cR))/c(\cR)$\footnote{Further details of our pre-processing, methodology and datasets are available in Section~\ref{sec:app-exp} of the appendix.}.

\subsection{Datasets}

We conduct experiments on two different datasets to show our methods work in different environments. 
The air quality dataset contains around $12000$ vertices and $16000$ edges. 
The Crucible dataset contains $260000$ vertices and $150000$ edges. 
These datasets are pre-procecessed as described above, thus reducing the size of the input graph before we execute our algorithms.

\textbf{Air quality in London:}
The goal is to minimise air pollution exposure of a runner in London such that the total distance of the route is within a given range.
We assume people run on the London road network. 
Vertices in $G$ represent road intersections and edges represent the roads. 
The weight of an edge is the distance of a road and the cost of an edge is the total pollution a runner is exposed to by running along the road.
The air quality model of London is a non-stationary mixture of Gaussian Processes \cite{Wilson:2012,Aglietti2019} that predicts air quality (nitrogen dioxide) from data such as sensors, road traffic and weather. 
We note that our methods are not dependent on the type of model used.
The output of the model is a two-dimensional grid which overlays the road network. 
The cost of an edge $e$ is the mean pollution of the grid squares intersecting $e$ in space multiplied by the weight of $e$. 
We assume the pollution (cost) is uniformly distributed along an edge.

\textbf{The Crucible:}
Our second application finds tours which seek a diverse variety of environment types in ``The Crucible'' map from the game Warcraft III \cite{Sturtevant2012}. 
The map consists of grid squares (512 by 512) belonging to one of the following five environment types (classes): normal ground (1), shallow water (2), trees (3), water (4) and out of bounds (5). 
An agent moving in the environment can only traverse normal ground, thus classes 2-5 trees are all defined as impassable for an agent. 
Let a vertex in $G$ represents a grid square in the map.
If a grid square is surrounded by a diverse variety of environment types, then the representative vertex will have high entropy.
We define the entropy of a vertex $v$ by $\mathcal{E}(v) = \sum_{c=1}^{C}~{- P_c \log P_c}$ where $C$ is the number of classes and $P_c$ is the probability of class $c$ appearing in a 7 by 7 grid centered on vertex $v$.
A unit weight edge $e_{u,v}$ exists between vertex $u$ and $v$ if both $u$ and $v$ are passable. 
The cost of an edge is defined by $c(e_{u,v}) = 1 - \frac{1}{2}(\mathcal{E}(u) + \mathcal{E}(v))$.

\subsection{Discussion} \label{sec:results}

\begin{table}[t]
  \caption{Mean overshoot (extra weight beyond $W_1$), mean margin of error (how far away is the heuristic to the applicable lower bound).
  Lower bounds (relaxations) are marked by $^*$.}
  
  \label{tab:results}
  \centering
  \begin{tabular}{l r r r r}
      \toprule
      & \multicolumn{2}{c}{Air Quality} & \multicolumn{2}{c}{The Crucible} \\
      \cmidrule(r){2-3} \cmidrule(r){4-5}
      Algorithm & Overshoot & $\%$ error & Overshoot & $\%$ error \\
      \midrule
      \rule{0pt}{1ex} \\
      Continuous relaxation (CR)$^*$ & 0.00 & 0.00 & 0.00 & 0.00 \\
      \deja heuristic (DjV) & 34.20 & 1.18 & 0.16 & 0.81 \\
      \rule{0pt}{1ex} \\
      Connectivity relaxation (XR)$^*$ & 17.16 & 0.00 & 0.00 & 0.00 \\
      Adaptive heuristic (AH) & 41.77 & 1.96 & 0.00 & 2.22 \\
      Suurballe's heuristic (SH) & 120.98 & 4.57 & 1.98 & 4.68 \\
      \bottomrule
  \end{tabular}
\end{table}

\begin{figure*}[t]
  \centering
  \includegraphics[width=\textwidth]{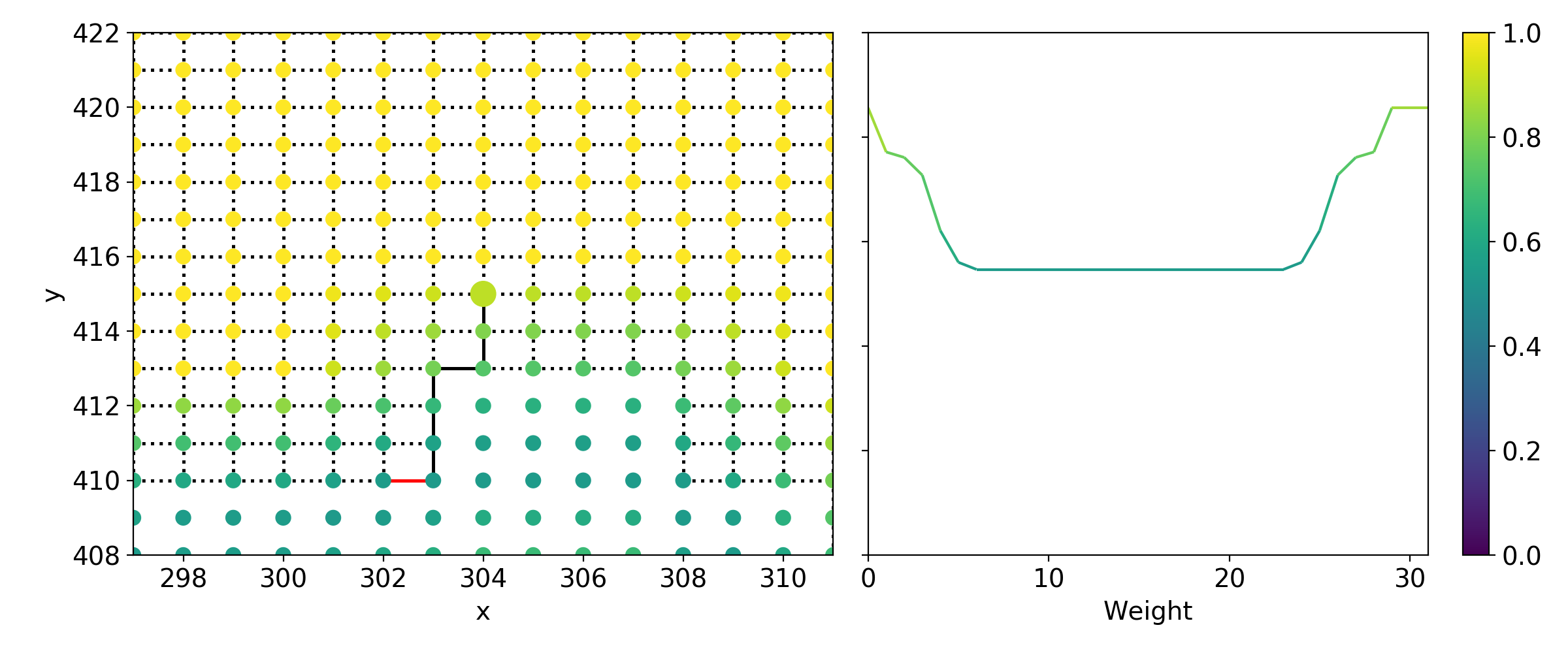}
  \caption{The \deja heuristic on the Crucible dataset for $W_1=30$, $W_2=35$.
  The colour scale represents the diversity of the environment: yellow is no diversity, dark blue is high diversity.
  [{\em Left}] The route taken. The origin is the large vertex at $x=304$, $y=415$. Solid edges in the tour are traversed exactly twice. The red edge is traversed 18 times. Dotted edges are not in the tour. The colour of a vertex represents $1-\cE(v)$.
  [{\em Right}] The cost of the route whilst traversing the tour. The colour of the line at a given weight is the diversity of the edge.}
  \label{fig:dejavu-crucible}
\end{figure*}

% the CLT problem

\textbf{CLT:} DjV yields consistently low-cost tours which have error $\approx 1\%$ on both datasets (Table~\ref{tab:results}) when compared to the CR lower bound.
Fig.~\ref{fig:dejavu-crucible} demonstrates the algorithm traversing a path to an edge with low-cost, repeating this edge 18 times (the long, flat line on right of Fig.~\ref{fig:dejavu-crucible}), before returning to the origin.
DjV traverses every edge with even multiplicity so the total weight of every tour will be even (assuming $w:E(G) \to \mathbb{N}$). Thus DjV does not consider low-cost solutions with odd weight, so if $W_1 = W_2$ and $W_1,W_2$ are {\em odd} integers, then DjV will not return a solution.
However, the $\cO(n \log n)$ time-complexity means DjV is fast (right of Figs. \ref{fig:pollution_dataset} and \ref{fig:map_dataset}) and the algorithm works well for a general $W_1, W_2$ in practise.

\begin{figure*}[t]
  \centering  
  \begin{subfigure}[t]{\textwidth}
    \includegraphics[width=\textwidth]{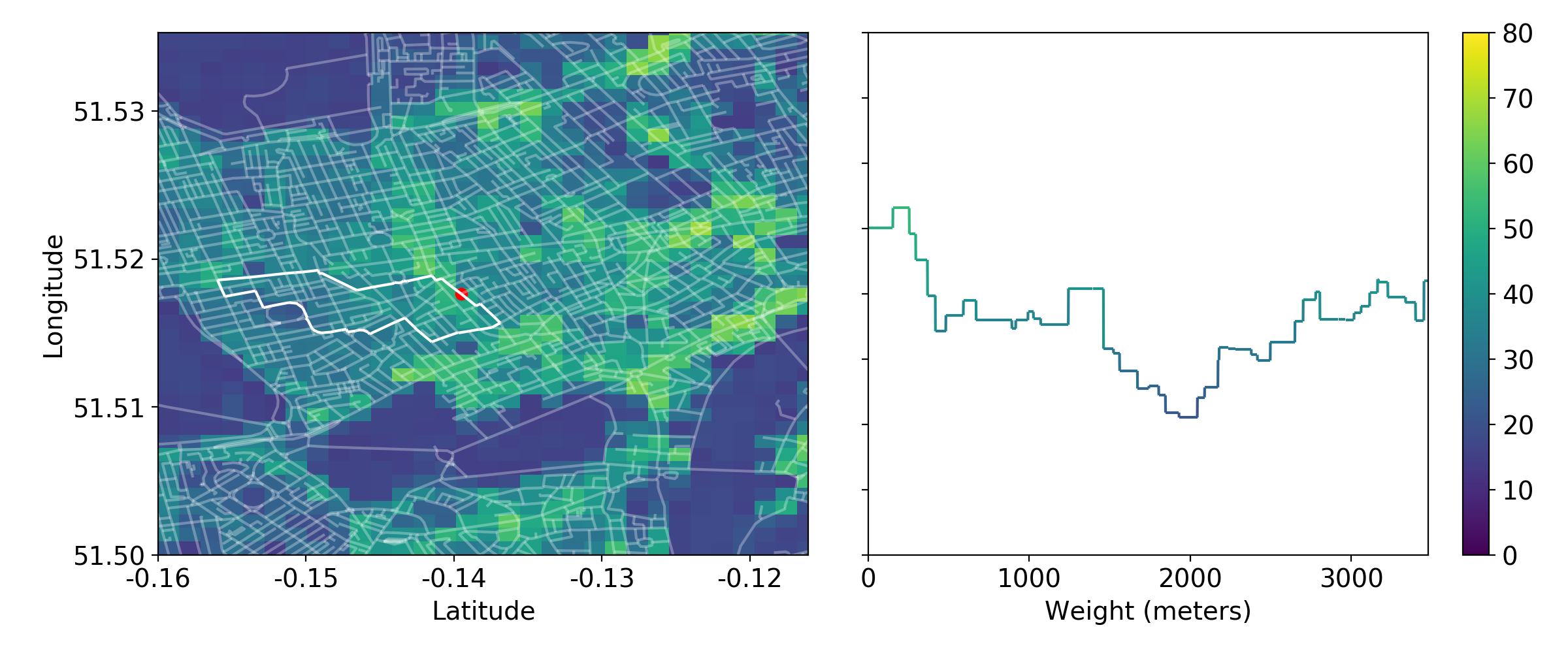}
  \end{subfigure}
  ~
  \begin{subfigure}[t]{\textwidth}
    \includegraphics[width=\textwidth]{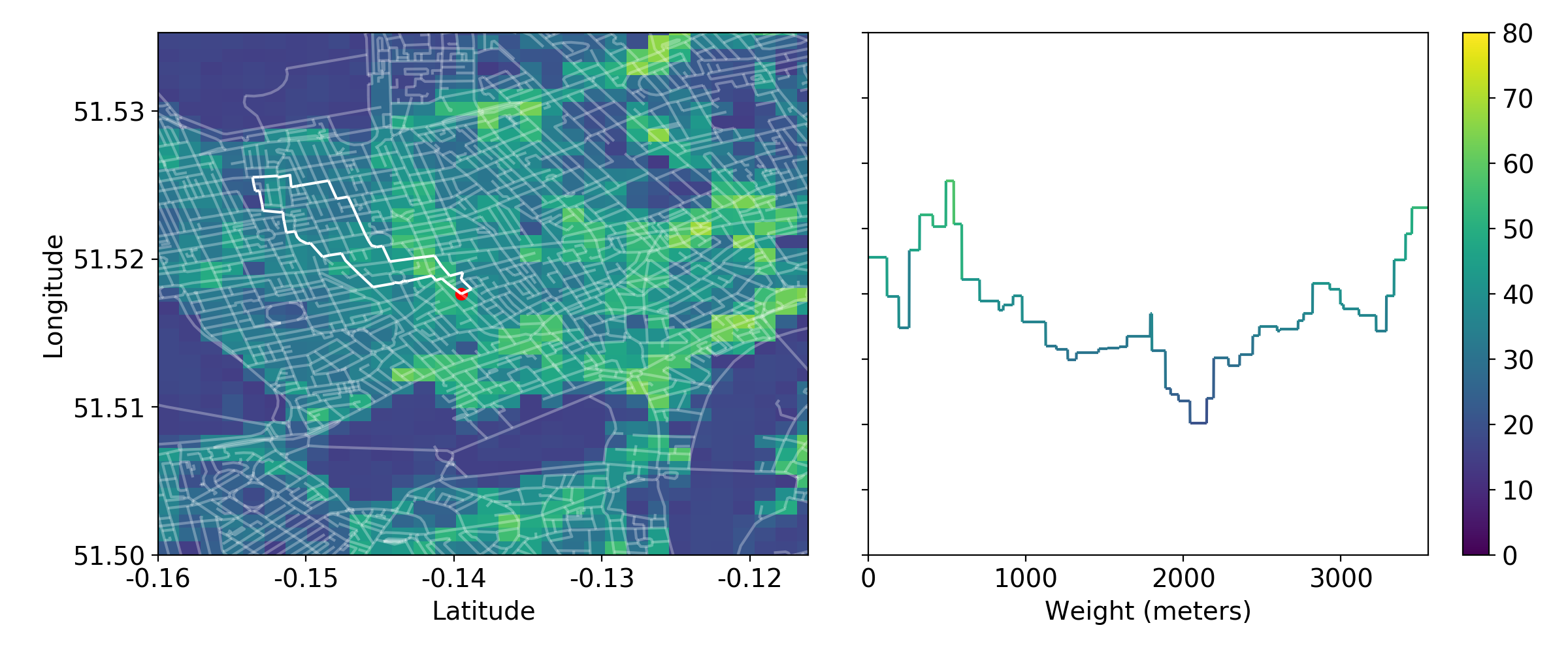}
  \end{subfigure}
  \caption{Routes computed by AH [{\em Top}] and SH [{\em Bottom}] starting at the same origin (red dot) for $W_1$ = 3.5km, $W_2$ = 3.75km. 
  [{\em Left}] Predicted nitrogen dioxide ($NO_2$) over central London measured in $\mu g m^{-3}$. 
  The colour scale is given on the right. 
  [{\em Right}] Air pollution exposure by running anti-clockwise around the route.}
  \label{fig:examples}
\end{figure*}

% the CLC problem

\textbf{CLC:} AH significantly outperforms SH on both datasets because it explores a larger proportion of the solution space.
% Fig.~\ref{fig:examples} shows a route computed by AH that explores more solutions, thus outperforming SH.
SH also significantly overshoots the lower weight threshold compared to AH (Table \ref{tab:results}), resulting in SH traversing more weight and thus (in general) more cost.
However, the trade-off for lower cost solutions is the $\cO(n \cdot m \log n)$ time-complexity of AH compared to the $\cO(m \log n)$ running time of SH.
The right of Figs. \ref{fig:pollution_dataset} and \ref{fig:map_dataset} clearly show this difference in time, and also highlights how long it takes for the integer program (IP) to optimally solve XR.
Indeed, the drop off in cost in Fig. \ref{fig:pollution_dataset} (left) for XR correlates with the point where the IP is no longer solving instances optimally because the running time is cut-off at 1 hour (Fig. \ref{fig:pollution_dataset} right).
In Fig. \ref{fig:map_dataset} (right), the IP quickly hits the cut-off time limit because the size of the Crucible graph is bigger, so the IP only returns a bound and not the optimal solution for $W_1 > 20$.
Thus for $W_1 > 2500m$ in the pollution dataset and $W_1 > 20$ for the Crucible dataset, the XR lower bound is actually larger than the given value, and so we are slightly over-estimating the error for AH and SH.

\section{Final remarks}

We have introduced the Constrained Least-cost Tour problem: an $\mathcal{NP}$-hard routing problem with the motivating application of finding running routes that minimise air pollution exposure in a city (see Fig.~\ref{fig:examples}).
We have derived relaxations and proposed heuristics for {\em weak} tours (CLT) and {\em strong} tours (CLC).
Experiments on both datasets show our algorithms perform competitively when compared to our derived lower bounds.
% Fig.~\ref{fig:examples} demonstrate our heuristics are effective on on real-world air quality predictions across the city of London.
Finally, the motivating application of ``running from air pollution" has a rich problem structure that we plan to further exploit; multiple pollutants, varying human sensitivities to different pollutants and uncertainty of the forecasting models. 

\section*{Acknowledgements}
Patrick O'Hara and Theodoros Damoulas are funded by the Lloyds Register Foundation programme on Data Centric Engineering through the London Air Quality project. 
This work was furthermore supported by The Alan Turing Institute for Data Science and AI under EPSRC grant EP/N510129/1 in collaboration with the Greater London Authority.
Patrick O'Hara was previously supported by the Warwick Impact Fund.
We would like to thank Oliver Hamelijnck (The Alan Turing Institute) for providing the air quality predictions of London and further thank the anonamous reviewers of IJCAI for their useful feedback.

\bibliographystyle{abbrvnat}
\bibliography{ref}

\newpage
\appendix

\section{$\NP$-hardness}

For completeness we give full proofs of Theorem 1, the $\NP$-hardness of CLC and the approximation of CLC.

\subsection{The CLT Problem}

We expand upon the proof of Theorem 1 in the paper. 
In particular, we expand on Claim 1. \\

\begin{theorem*}
     The Constrained Least-cost Tour Problem is $\NP$-hard, even when the input graph is a path.
\end{theorem*}
\begin{proof}
  Recall from the paper that we construct an instance of {\sc CLT Decision} from an instance $I$ of {\sc Subset Sum}.
  We argue that $I$ is a {\sc Yes}-instance of {\sc Subset Sum} if and only if the constructed instance is a {\sc Yes}-instance of {\sc CLT Decision}.
  
  Suppose that $I$ is a {\sc Yes}-instance of {\sc Subset Sum} and let $(y_1,\dots, y_n)$ be the solution. Consider the ``natural" tour $\mathcal{T}$ in $G$ starting at $v_1$.
  For every $i \in [n]$, $\cT$ traverses $e_i$ exactly $2y_i+2$ times.
  The edge $e_{n+1}$ is traversed exactly twice. 
  Thus, the value of $w(\mathcal{T})$ is precisely the following:
  
  \begin{align*}
  w(\mathcal{T}) &= \sum_{i\in [n]}~w(e_i)(2y_i+2)+2w(e_{n+1}) \\
  &=2\sum_{i \in [n]}~{y_i \cdot w(e_i)} + 2\sum_{i \in [n]}~{w(e_i)} + 2 w(e_{n+1}) \\
  &=2\overbrace{\sum_{i\in [n]}~{y_i \cdot f(z_i)}}^{Z}+2\sum_{i\in[n]}~w(e_i) + 2 w(e_{n+1})=W_1
  \end{align*}
  
  Similarly the value of $c(\mathcal{T})$ is:
  
  \begin{align*}
        c(\mathcal{T}) &= \sum_{i\in [n]}~c(e_i)(2y_i+2)+2c(e_{n+1})\\
        &=2\sum_{i \in [n]}~{y_i \cdot w(e_i)} + 2\sum_{i \in [n]}~{w(e_i)} + \overbrace{2Z + 2\sum_{i\in[n]}~w(e_i)}^{2c(e_{n+1})} \\
        &=2\overbrace{\sum_{i\in [n]}~y_i\cdot f(z_i)}^{Z}+2\sum_{i\in[n]}~w(e_i) + 2Z + 2\sum_{i\in[n]}~w(e_i) =C
    \end{align*}
  Hence, we conclude that if $I$ is a {\sc Yes}-instance, then the constructed instance of {\sc CLT Decision} is a {\sc Yes}-instance. 
  
  Conversely, suppose that the constructed instance of {\sc CLT Decision} is a {\sc Yes} instance. 
  This implies we have a tour $\mathcal{T}$ in $G$ starting at $v_1$ such that $W_1 \leq w(\mathcal{T})\leq W_2$ and $c(\mathcal{T})\leq C$.

\begin{claim*}
$\mathcal{T}$ traverses every edge of $G$ at least twice and the edge $e_{n+1}$ exactly twice.
\end{claim*}

\begin{proof}
    Observe that in order to prove the claim, it is sufficient to prove that $\mathcal{T}$ traverses $e_{n+1}$ {\em exactly twice}. 
    This is because $G$ is a path and any tour containing $v_1$ and $v_{n+2}$ must traverse every edge of $G$ at least once and any {\em tour} in $G$  containing $v_1$ must traverse every edge of $G$ an even number of times. 
    Consequently, we now focus on proving that $\mathcal{T}$ traverses $e_{n+1}$ exactly twice.

    We first show that $\mathcal{T}$ traverses $e_{n+1}$ {\em at least} twice. 
    Suppose not. 
    Then, $\mathcal{T}$ does not traverse $e_{n+1}$ at all. 
    Note that we have defined $W_1$ such that $W_1\geq 2w(e_{n+1})$ and since $\mathcal{T}$ is a solution for the instance of {\sc CLT Decision}, we know that $w(\mathcal{T})\geq W_1$. 
    Then, by the pigeonhole principle, there must be an edge $e_i\in \{e_1,\dots, e_n\}$ such that the multiplicity of $e_i$ in $\mathcal{T}$ is greater than $W_1/w(e_i)$ and hence greater than $4Z+4\sum_{i\in [n]}w(e_i)$, which is precisely the value of $C$. 
    This contradicts our assumption that $c(\mathcal{T})\leq C$. Hence we conclude that $\mathcal{T}$ traverses $e_{n+1}$ at least twice. 

    It remains to argue that $\mathcal{T}$ traverses $e_{n+1}$ {\em at most} twice. Suppose that this is not the case and that the edge $e_{n+1}$ occurs at least 3 times in $\mathcal{T}$. Then, it must be the case that $e_{n+1}$ appears at least 4 times in $\mathcal{T}$. In this case, $c(\mathcal{T})>4\cdot c(e_{n+1})$ and $c(\mathcal{T})\leq C$, which is a contradiction since we chose $c(e_{n+1})$ in such a way that $C=4 c(e_{n+1})$. Hence we conclude that $\mathcal{T}$ traverses $e_{n+1}$ at most twice.
%  This completes the proof of the claim. 
\end{proof}

    We now describe the solution vector $(y_1,\dots, y_n)$ for the {\sc Subset Sum} instance.
    Recall we have already proved that for all $e\in E(G)$: $\phi(\mathcal{T},e)\geq 2$ and $\phi(\mathcal{T},e)$ is even.
    For the last edge: $\phi(\mathcal{T},e_{n+1})=2$.
    For every $i\in [n]$, define $y_i=\frac{1}{2} \phi(\mathcal{T},e_i)-2.$ 
    Clearly, $y_i\in {\mathbb N}\cup \{0\}$ for all $i\in [n]$ as required in the description of {\sc Subset Sum}.
    
    It remains to argue that $\sum_{i \in [n]}~{y_i \cdot f(z_i)} = Z$.
    Since $\mathcal{T}$ traverses $e_{n+1}$ exactly twice, we first use the fact that $w(\mathcal{T})\geq W_1$ to conclude that $\sum_{i\in [n]} y_i\cdot f(z_i)\geq Z$:
    
    \begin{align*}
        w(\cT) &\geq W_1 \\
        2w(e_{n+1}) + \sum_{i \in [n]}~{\phi(\cT,e_i)\cdot w(e_i)} &\geq 2Z + 2w(e_{n+1}) + 2\sum_{i \in [n]}~{w(e_i)} \\
        \sum_{i \in [n]}~{2(y_i + 2)w(e_i)} &\geq 2Z + 2\sum_{i \in [n]}~{w(e_i)} \\
        \sum_{i \in [n]}~{y_i \cdot f(z_i)} &\geq Z
    \end{align*}
    
    and use $c(\mathcal{T})\leq C$ to infer that $\sum_{i\in [n]} y_i\cdot w(z_i)\leq Z$:
     \begin{align*}
         c(\cT) &\leq C \\
         2c(e_{n+1}) + \sum_{i \in [n]}~{\phi(\cT,e_i)c(e_i)} &\leq 4Z + 4\sum_{i \in [n]}~{w(e_i)} \\
         2Z + 2\sum_{i \in [n]}~{w(e_i)} + \sum_{i \in [n]}~{2(y_i + 2)w(e_i)} &\leq 4Z + 4\sum_{i \in [n]}~{w(e_i)} \\
         2\sum_{i \in [n]}~{y_i \cdot w(e_i)} + 2\sum_{i \in [n]}~{w(e_i)} &\leq 2Z + 2\sum_{i \in [n]}~{w(e_i)} \\
         \sum_{i \in [n]}~{y_i \cdot f(e_i)} &\leq Z
     \end{align*}
    
    This completes the proof in the converse direction, and thus the proof of the theorem.
  \end{proof}

\subsection{The CLC Problem}

\begin{definition*}
    Given an undirected graph $G$ with vertices $V(G)$ and edges $E(G)$, the Hamiltonian Cycle (HC) problem is to find a cycle $\cC$ that visits every vertex in the graph exactly once.
\end{definition*}

Given a graph $G$ with a weight function $w:E(G) \to \mathbb{N}$ and cost function $c:E(G) \to \mathbb{N}$ on the edges; a start vertex $\origin \in V(G)$; two weight thresholds $W_1,W_2 \in \mathbb{N}$ and a cost threshold $C \in \mathbb{N}$, the {\sc CLC-Decision} problem asks is there a strong tour $\cT$ starting and ending at $\origin$ such that $W_1 \leq w(\cT) \leq W_2$ and $c(\cT) = C$. \\

\begin{theorem*}
    The Constrained Least-cost Cycle (CLC) Problem is $\NP$-hard.
\end{theorem*}

\begin{prf}
    To prove {\sc CLC-Decision} is $\NP$-hard, we find a polynomial-time reduction from HC.
    We note that HC is $\NP$-complete.
    Let $I$ be an instance of HC on a graph $G$ with vertices $V(G) = \{v_1, \ldots, v_n\}$ and edges $E(G)$.
    We construct an instance of {\sc CLC-Decision} as follows.
    Let the cost and the weight of every edge be 1.
    Set $W_1 = W_2 = C = n$ where $n$ is the number of vertices in $G$.
    Let $s^\star = v_1$.
    The reduction is clearly polynomial.
    We prove that $I$ is a {\sc Yes}-instance of HC if and only the constructed instance of {\sc CLC-Decision} is a {\sc Yes}-instance.
    
    Suppose $I$ is a {\sc Yes}-instance.
    Then $I$ is a simple cycle $\cC$ that visits every vertex in the graph (including $v_1$) exactly once.
    The cycle is a strong tour $\cT$ that traverses exactly $n$ edges and includes the start vertex $v_1$.
    Thus the total weight of $\cT$ on the constructed instance of {\sc CLC-Decision} is $w(\cT) = n = W_1 = W_2$ and the total cost is $c(\cT) = n = C$. 
    Hence we conclude the constructed instance is a {\sc Yes}-instance.
    
    Now suppose the constructed instance of {\sc CLC-Decision} is a {\sc Yes}-instance.
    Then we have a strong tour $\cT$ that starts at the origin $v_1$, visits every vertex {\em at most once}, and has total weight $n = W_1 = W_2$ and total cost $n = C$.
    We need to show that $\cT$ visits every vertex exactly once.
    Suppose otherwise.
    Then there must exist a vertex $u$ that has not been visited by $\cT$, which implies $\cT$ is a simple cycle on at most $n-1$ vertices that has total weight and total cost of $n$.
    This is clearly not possible since edges have unit weight and unit cost.
    Thus we conclude that $\cT$ visits every vertex exactly once and the strong tour is a Hamiltonian cycle.
\end{prf}

\subsection{Approximation of CLC}

\begin{definition*}
    Let OPT be the cost of the optimal solution to a problem and $\alpha \in {\mathbb R}$ be greater than or equal to 1. An algorithm is an $\alpha$-approximation algorithm if and only if for every instance of the problem it returns a solution within a factor $\alpha$ of OPT.
\end{definition*}

When referring to an $\alpha$-approximation algorithm, we shall mean that the algorithm must run in polynomial-time.

\begin{lemma*}
    For CLC without the triangle inequality assumption, there does not exist an $\alpha$-approximation algorithm for any $\alpha \geq 1$, provided ${\cal P} \neq \NP$.
\end{lemma*}

\begin{prf}
    Given an instance $G$ of HC with vertices $V(G)$ and edges $E(G)$, construct an instance of CLC on a complete graph $\cK$ as follows. 
    For all $e \in E(G)$, let $c(e) = w(e) = 1$ in $\cK$. 
    For all pairs of vertices $i,j \in V(G)$ for which $e_{ij} \notin E(G)$ and $i \neq j$, let $c(e_{ij}) = \alpha \cdot n$ and $w(e_{ij}) = 1$.
    Let $W_1 = W_2 = n$ where $n$ is the number of vertices.
    Assume there exists an $\alpha$-approximation algorithm ({\sc Apx}) for CLC.
    We show that such an algorithm can be used to solve HC in polynomial time.

    First suppose there exists a Hamiltonian cycle in $G$.
    Then the optimal solution OPT for CLC will have cost $n$ and weight $n$, so $c(\text{\sc Apx}) \leq \alpha \cdot n$. Now suppose there does not exists a Hamiltonian cycle in $G$.
    Then OPT must use one edge not in $G$ with cost $\alpha \cdot n$.
    The cost of OPT will therefore be $c(OPT) \geq n-1 + \alpha \cdot n$.
    
    Hence we conclude that $G$ has a Hamiltonian cycle if and only if the cost of {\sc Apx} is at most $\alpha \cdot n$.
\end{prf}

\section{An Integer Programming Formulation of the CLC Problem}

Recall that $x_{ij}$ is a 0-1 variable placed on the edges in graph $G$, $\cA_i$ is the set of edges adjacent to vertex $i$, and $x(S)$ is the sum of $x_{ij}$ variables on edge set $S \subset E(G)$.

\begin{align}
    & \text{min} & & \displaystyle\sum_{e_{ij} \in E(G)}~{x_{ij} \cdot c(e_{ij})} & \\
    & \text{s.t.} & & W_1 \leq \displaystyle\sum_{e_{ij} \in E(G)}~{x_{ij} \cdot w(e_{ij})} \leq W_2 & \\
    & & & x(\cA_{\origin}) = 2 & \\
    & & & \frac{1}{2} x(\cA_i) \in \{0,1\} & \forall i \in V(G) \\
    & & & x_{ij} \in \{0,1\} & \forall e_{ij} \in E(G) \\
    & & & \text{Subtour elimination constraints} & 
\end{align}

There are several ways to define the subtour elimination constraint.
We give one such formulation using cutsets:
	
	$$\forall S \in \Phi(i,j), \forall i,j \in V: x(S) \geq x(\mathcal{A}_i) + x(\mathcal{A}_j) - 2$$
	
where $\Phi(i,j)$ is the set of minimal $ij$-edge cuts. 
That is, each cut in $\Phi(i,j)$ is a minimal set of edges that if removed would disconnect $i$ and $j$ in the graph.
The connectivity relaxation uses the objective function (1) with contraints (2)-(5) from the IP formulation above, but relaxes constraint (6).

\section{Suurballe's Algorithm}

In the CLT/CLC problem, we are given an undirected graph $G$ with vertices $V(G)$ and edges $E(G)$.
Edges have a weight function $w:E(G) \to \mathbb{N}$ and a cost function $c:E(G) \to \mathbb{N}$.
However, Suurballe's algorithm requires a directed input graph that is asymmetric, that is if $(x,y)$ is an arc in the graph, then $(y,x)$ is not in the graph.

We construct a directed, asymmetric graph $G^\prime$ from the undirected graph $G$ as follows.
For each vertex $v \in V(G)$, split $v$ into two vertices $v_1$ and $v_2$ and add them to $V(G^\prime)$. 
Add a directed \textit{split} arc from $v_1$ to $v_2$ in $G^\prime$ with zero cost and zero weight. 
For every undirected edge $e_{u,v}$ adjacent to $v$ in $G$, add a directed arc from $v_2$ to $u_1$ in $G^\prime$ with the same weight and cost as $e_{u,v}$ in $G$. 
The construction requires $O(m + n)$ time and space complexity.

\section{Experiments} \label{sec:app-exp}

In this section, we expand upon the pre-processsing algorithms, methodology and datasets uses for the experiments. To run our algorithms on the same datasets, please refer to our GitHub repository\footnote{\url{https://patrickohara.github.io/CLT-problem/}}.

\subsection{Pre-processing}

We use two pre-processing methods to reduce the size of the graph.
The time taken for pre-processing is not included when timing the algorithms on the right of Figures \ref{fig:pollution_dataset} and \ref{fig:map_dataset}.
The first pre-processing algorithm removes vertices from the graph that cannot be reached from the origin within weight $\frac{W_2}{2}$, that is, if the weight of the shortest path from the origin to a vertex $u$ is greater than $\frac{W_2}{2}$, then remove $u$ from the graph (see Algorithm \ref{alg:prune}).
The second removes vertices with degree one (leaves) and is given in Algorithm \ref{alg:leaves}.
It uses a recursive depth first search to remove leaves from the graph.

\begin{algorithm}[h]
    \DontPrintSemicolon
    \SetKwInOut{Input}{input}\SetKwInOut{Output}{output}
    \caption{Prune unreachable vertices from the graph.}
    \label{alg:prune}
    \SetAlgoLined
    \Input{An undirected graph $G$ with edge weights $w:E(G) \to \mathbb{N}$ and edge costs $c:E(G) \to \mathbb{N}$; the origin $\origin \in V(G)$; upper weight thresholds $W_2 \in \mathbb{N}$.}

    Construct a shortest path tree $T$ rooted at $\origin$ with respect to weight using Dijkstra's algorithm. \;
    Store the parent $\pi_u$ and length $l_u$ of the shortest path w.r.t. weight for every vertex $u$. \;
    For every vertex $u \in V(G)$, if $l_u > \frac{W_2}{2}$, then remove $u$ from $G$. \;
    \Output{A pruned graph.}
\end{algorithm}

\begin{algorithm}[h]
    \DontPrintSemicolon
    \SetKwInOut{Input}{input}\SetKwInOut{Output}{output}
    \SetKwFunction{proc}{proc}
    \caption{Remove leaves from the input graph.}
    \label{alg:leaves}
    \Input{An undirected graph $G$; the origin $\origin \in V(G)$.}
    Start depth first search from $\origin$ \;
    If the current vertex $u$ has degree one, then remove $u$ from $G$. \;
    Recurse up the tree removing vertices with degree one. \;
    \Output{A graph with no leaves.}
\end{algorithm}

\subsection{Datasets}

We compare our heuristics on two different datasets to show that our methods can be applied in different contexts.
The contrast between the air quality (AQ) dataset and the Crucible dataset is also interesting.
Firstly, the structure of the graph in the Crucible is a grid, whereas the AQ graph is a road network.
Running heuristics on different types of graphs can often highlight strengths of weaknesses, although in our experiments there were no such notable strengths or weaknesses.
Secondly, the cost function on edges in the two datasets is spatially distributed in a very different way. 
In the Crucible, there are large areas of space where in all edges will have uniform cost.
Compare this to AQ in which there are large areas of space with relatively low air pollution (cost) with localised peaks of highly polluted air.
Despite these differences, our heuristics have a similar margin of error on both datasets when compared to their respective relaxations (Table~\ref{tab:results}).

\begin{figure}[t]
  \centering
  \includegraphics[width=\textwidth]{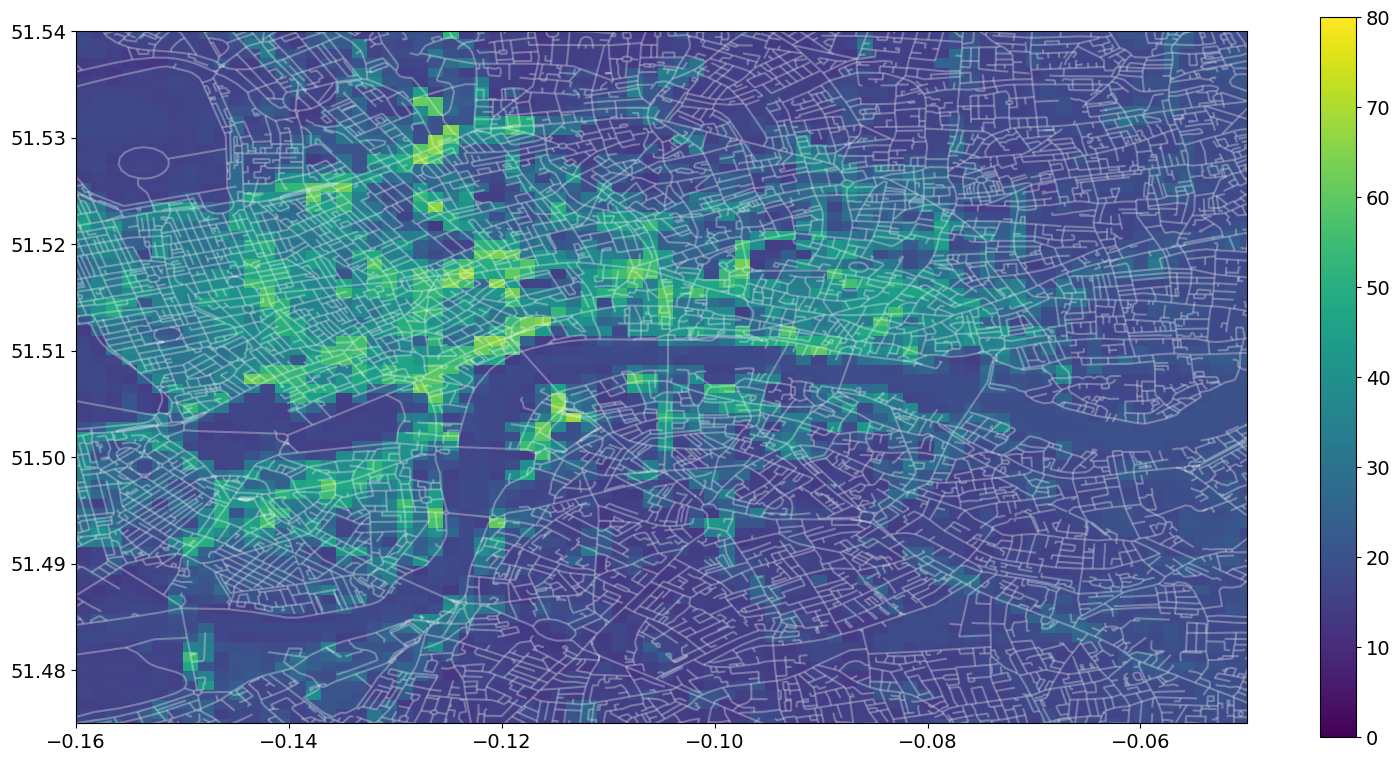}
  \caption{Snapshot from the air quality model of London. The prediction is for nitrogen dioxide ($NO_2$). $NO_2$ is measured in $\mu g m ^{-3}$. Yellow shows high air pollution, dark blue is low air pollution.}
  \label{fig:air-quality}
\end{figure}

\paragraph{Air quality in London} The road network of London was downloaded from the Ordnance Survey\footnote{\url{https://www.ordnancesurvey.co.uk}}. 
The air quality prediction shown in Fig. \ref{fig:air-quality} are a snapshot from an AQ model of London.
The model is currently under development at the Alan Turing Institute.
We emphasise that our algorithms are not dependent upon the model of air quality.
The road network was pruned using an SQL PostGIS query to return all roads that intersect with the prediction area of the AQ model.

\begin{figure*}[t]
    \centering      
    \begin{subfigure}[t]{0.45\textwidth}
        \includegraphics[width=\linewidth]{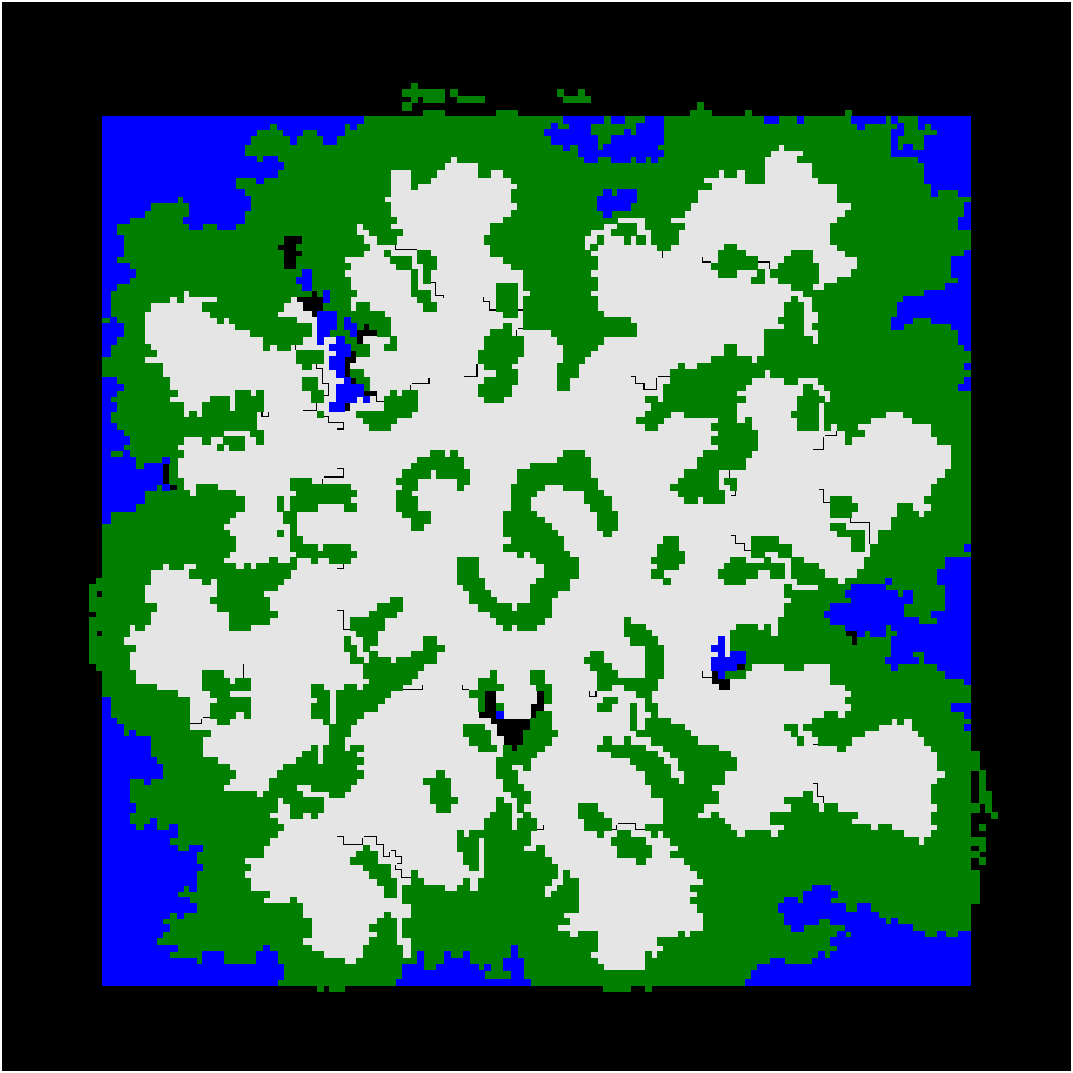}
        \caption{The original dataset showing different types of environment. Black represents impassible, blue shows water, green shows trees and white shows passible ground.}
        \label{fig:map}
    \end{subfigure}
    ~
    \begin{subfigure}[t]{0.5\textwidth}
      \centering
       \includegraphics[width=\linewidth]{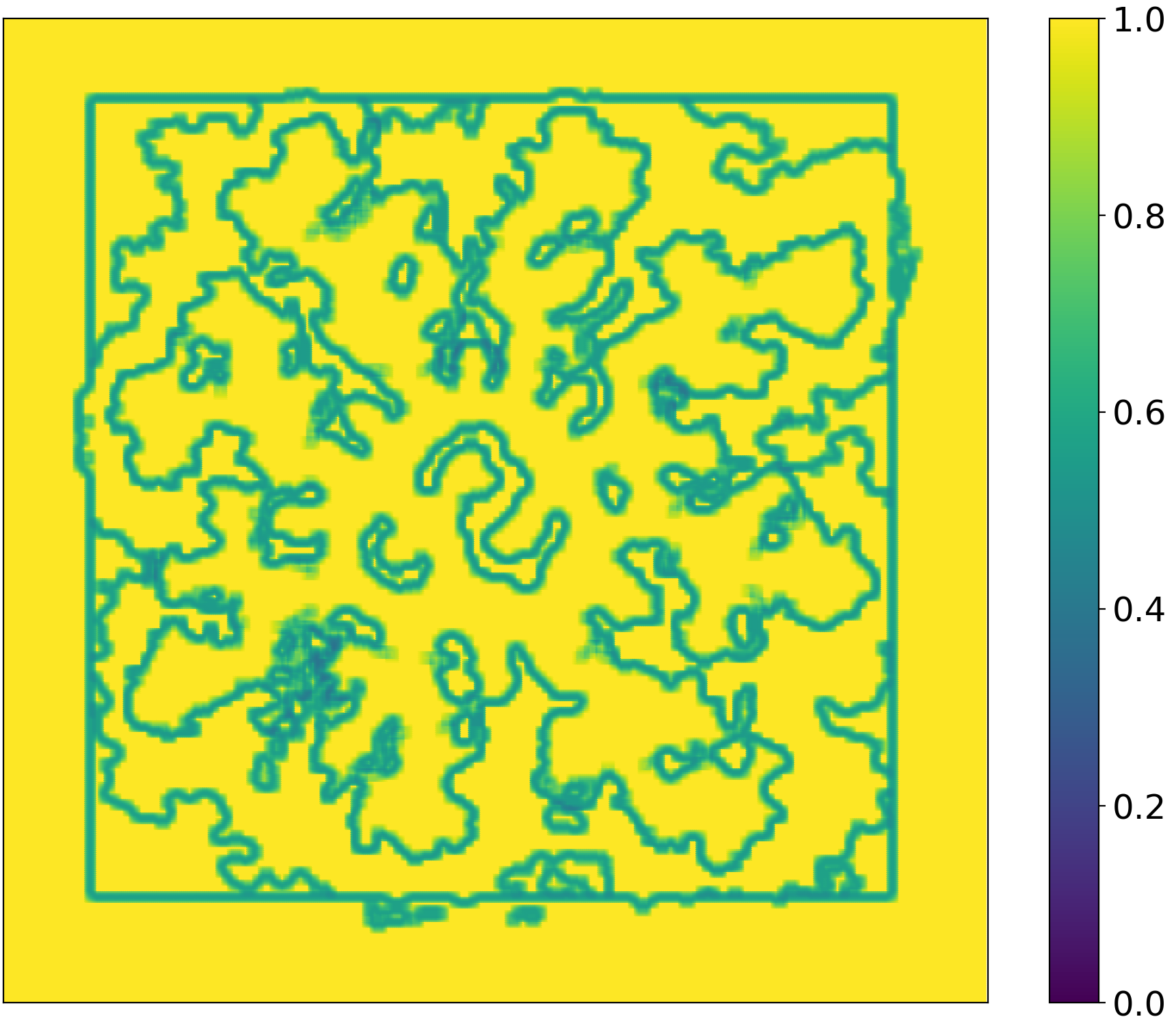}
       \caption{Areas with high diversity (dark blue) and low diversity (yellow) of environment types. This quantity is calculated by $1 - \cE(v)$.}
       \label{fig:borders}
    \end{subfigure}
    \caption{Visual representation of how the diversity is computed for the Crucible dataset. Note the structure of dark areas in (b) which follows the borders between environment types in (a).}
    \label{fig:crucible}
\end{figure*}

\paragraph{The Crucible} Fig.~\ref{fig:crucible} shows the Crucible dataset which can be downloaded from Moving AI\footnote{\url{https://www.movingai.com/benchmarks/wc3maps512/index.html}}.
The aim of the agent is to find a tour that visits a diverse range of environment types.
The original dataset is shown in Fig.~\ref{fig:map} and the diversity of environment types is shown in Fig.~\ref{fig:borders}.
The agent will seek the darker areas of Fig.~\ref{fig:borders}.
The darker areas show borders between different environment types, for example where the normal ground (class 1) meets trees (class 3).

\subsection{Additional details}

The heuristics and relaxations are coded using Python.
Specifically, we use the {\tt networkx} library to store the graph datastucture.
The machines used to compute the results are given in the paper: the first was an Azure virtual machine (VM) for the heuristics and the continuous relaxation (CR); and the second was an IBM Cloud machine running the CPLEX library for the connectivity relaxation (XR).
Using a more powerful machine for the XR means that the time taken to compute a result is not comparible to the time taken to compute a result for heuristics and CR.
However, Figs.~\ref{fig:pollution_dataset} and \ref{fig:map_dataset} [right] show that even with a more powerful machine, XR still takes substantially more time to find an solution that any of the heuristics or CR.
Further, the important algorithms to compare in terms of time are the Adaptive heuristic and Suurballe's heuristic, since they are the two competing algorithms for the CLC problem.

For each dataset, 10 vertices were chosen randomly from a uniform distribution.
Each dataset was given 10 weight thresholds 
Each heuristic and each relaxation were tested for each configuration.
For each solution to a relaxation, we record the total weight, total cost, number of vertices in the solution, the overshoot, number of vertices and edges in the pre-processed input graph and the time taken in seconds.
In addition to the above quantities, we calculate the margin of error for each heuristic compared to the appropriate relaxation.

\end{document}